\documentclass[12pt]{article}
\usepackage{epsf}
\def\beq{\begin{equation}}
\def\eeq{\end{equation}}
\def\bea{\begin{eqnarray}}
\def\eea{\end{eqnarray}}

\def\vel{\left|}
\def\ver{\right|}

\def\ga{\left(}
\def\dr{\right)}

\def\la{\langle}
\def\ra{\rangle}
\def\ba{\begin{array}}
\def\ea{\end{array}}

\title{ {\bf
The CP asymmetry for $B\rightarrow K^* l^+ l^-$ decay in the general two 
Higgs doublet model}}
\author{\vspace{1cm}\\
        {\bf E. O. Iltan}
        \thanks{E-mail address:
        eiltan@heraklit.physics.metu.edu.tr}
 \\
        Physics Department, Middle East Technical University \\
        Ankara, Turkey\\}

\date{}

\begin{document}
\setlength{\baselineskip}{24pt}
\maketitle
\setlength{\baselineskip}{7mm}
\begin{abstract}
We study the CP asymmetry for the exclusive decay 
$B\rightarrow K^* l^+ l^-$ in the two Higgs doublet model with three level
flavor changing neutral currents (model III). We analyse the dependency of 
this quantity  to the new phase coming from the complex Yukawa couplings in
the theory and we find that there exist a considerable CP violation  
for the relevant process. Further, we see that the sign of the
Wilson coefficient $C_7^{eff}$ can be determined by fixing dilepton mass. 
Therefore, the future measurements of the CP asymmetry for
$B\rightarrow K^* l^+ l^-$ decay will give a powerful information about the
sign of Wilson coefficient $C_{7}^{eff}$ and new physics beyond the SM.
\end{abstract} 
\thispagestyle{empty}
\newpage
\setcounter{page}{1}

\section{Introduction}
Rare B-decays are induced by flavor changing neutral currents 
(FCNC) at loop level in the Standard model (SM). Therefore,
the measurements of the physical quantities, like Branching ratio (Br), CP
asymmetry ($A_{CP}$), forward backward asymmetry (AFB), in such decays, 
provide a powerful test for the SM and they give a comprehensive 
information about the fundamental parameters, such as Cabbibo-Kobayashi-
Maskawa (CKM) matrix elements, leptonic decay constants, etc. Further, 
they play an important role in the determination of the
physics beyond the SM, such as two Higgs Doublet model (2HDM), Minimal 
Supersymmetric extension of the SM (MSSM) \cite{Hewett}, etc. 
With the measurement of the Branching ratios ($Br$)
of the inclusive $B\rightarrow X_s\gamma$ \cite{cleo} and the exclusive
$B\rightarrow K^*\gamma$ \cite{rammar} decays, the studies on rare B 
decays have been increased. 

Among rare B decays, $B\rightarrow K^*l^+ l^-$ decay, induced by the
inclusive process $b\rightarrow s l^+ l^-$, becomes attractive since it has
a large $Br$ in the framework of the SM and it can be measured in future
experiments. In the literature, these decays have been studied in the SM, 
2HDM and MSSM  \cite{R4}- \cite{alil3} extensively.
For $b\rightarrow s l^+ l^-$ induced processes, the matrix element contains
a term proportional to $V_{tb}V_{ts}^*$, $V_{cb}V_{cs}^*$  and
 $V_{ub}V_{us}^*$ coming from $t\bar{t}$, $c\bar{c}$ and $u\bar{u}$
quark loops respectively. The unitarity of CKM, $V_{ib}V_{is}^*=0\,\,
, (i=u,c,t)$, causes that this term is only proportional to $V_{tb}V_{ts}^*$ 
since $V_{ub}V_{us}^*$ is smaller compared to $V_{tb}V_{ts}^*$. 
Therefore, CP violating effects are
suppressed in the SM. However, there is a new source for CP violation
in the framework of the general 2HDM, so called model III. In this model,
extra phase angles can appear in the Yukawa couplings when they are taken
complex. In \cite{wolfenstein}, the effect of the phase to the decay 
$b\rightarrow s\gamma$ was studied. Recently, the constraints on the phase 
angle in the product of Yukawa coupligs $\lambda_{bb}\lambda_{tt}$ was 
predicted by \cite{david}. These angles can cause an observable CP 
violation in the $b\rightarrow  s l^+ l^-$ induced decays. The theoretical 
investigation of CP violation effects in the model III for 
$b\rightarrow  s l^+ l^-$ and its induced exclusive decays, such as 
$B\rightarrow  K^* l^+ l^-$, can be an important test for the new physics, 
since there is almost no such effect in the context of the SM. 

Even if the theoretical analysis of exclusive decays is more complicated 
due to the hadronic form factors, the experimental investigation of them  
is  easier compared those of inclusive ones. Therefore, in this work, we 
study the CP violating effects in the model III for the exclusive 
$B\rightarrow K^* l^+ l^-$ decay.

The paper is organized as follows:
In Section 2, we present the matrix element for the inclusive 
$b\rightarrow s l^+ l^- \,\, (l=e,\mu)$ decay and calculate $A_{CP}$ 
in the framework of the model III. Section 3 is devoted to discussion and 
our conclusions.
\section{CP violation in the exclusive decay $B\rightarrow K^* l^+ l^-$ in
the framework of the model III} 
Before starting with the exclusive decay $B\rightarrow K^* l^+ l^-$ 
($l=e,\mu$), we would like to give a brief summary about the model III 
and to derive the matrix element of the inclusive decay 
$b\rightarrow s l^+ l^-$ which induces the exclusive 
$B\rightarrow K^* l^+ l^-$ process. 

In the general 2HDM , called model III, 
the Yukawa interaction can be defined as 
\begin{eqnarray}
{\cal{L}}_{Y}=\eta^{U}_{ij} \bar{Q}_{i L} \tilde{\phi_{1}} U_{j R}+
\eta^{D}_{ij} \bar{Q}_{i L} \phi_{1} D_{j R}+
\xi^{U}_{ij} \bar{Q}_{i L} \tilde{\phi_{2}} U_{j R}+
\xi^{D}_{ij} \bar{Q}_{i L} \phi_{2} D_{j R} + h.c. \,\,\, ,
\label{lagrangian}
\end{eqnarray}
where $L$ and $R$ denote chiral projections $L(R)=1/2(1\mp \gamma_5)$,
$\phi_{i}$ for $i=1,2$, are the two scalar doublets. The Yukawa matrices  
$\eta^{U,D}_{ij}$ and $\xi^{U,D}_{ij}$ have  in general complex entries.
With the choice of $\phi_{1}$ and $\phi_{2}$,
\begin{eqnarray}
\phi_{1}=\frac{1}{\sqrt{2}}\left[\left(\begin{array}{c c} 
0\\v+H^{0}\end{array}\right)\; + \left(\begin{array}{c c} 
\sqrt{2} \chi^{+}\\ i \chi^{0}\end{array}\right) \right]\, ; 
\phi_{2}=\frac{1}{\sqrt{2}}\left(\begin{array}{c c} 
\sqrt{2} H^{+}\\ H_1+i H_2 \end{array}\right) \,\, .
\label{choice}
\end{eqnarray}
and the vacuum expectation values,  
\begin{eqnarray}
<\phi_{1}>=\frac{1}{\sqrt{2}}\left(\begin{array}{c c} 
0\\v\end{array}\right) \,  \, ; 
<\phi_{2}>=0 \,\, ,
\label{choice2}
\end{eqnarray}
it is possible to collect SM particles in the first doublet and new
particles in the second one. The Flavor Changing (FC) part of the
interaction can be written as 
\begin{eqnarray}
{\cal{L}}_{Y,FC}=
\xi^{U}_{ij} \bar{Q}_{i L} \tilde{\phi_{2}} U_{j R}+
\xi^{D}_{ij} \bar{Q}_{i L} \phi_{2} D_{j R} + h.c. \,\, ,
\label{lagrangianFC}
\end{eqnarray}
where the couplings  $\xi^{U,D}$ for the FC charged interactions are
\begin{eqnarray}
\xi^{U}_{ch}&=& \xi_{N} \,\, V_{CKM} \nonumber \,\, ,\\
\xi^{D}_{ch}&=& V_{CKM} \,\, \xi_{N} \,\, ,
\label{ksi1} 
\end{eqnarray}
and $\xi^{U,D}_{N}$ is defined by the expression (more details see 
\cite{soni})
\begin{eqnarray}
\xi^{U,D}_{N}=(V_L^{U,D})^{-1} \xi^{U,D} V_R^{U,D}\,\, .
\label{ksineut}
\end{eqnarray}
Note that the index "N" in $\xi^{U,D}_{N}$ denotes the word "neutral". 

The procedure is to obtain the effective Hamiltonian and calculate the QCD
corrections by matching the full theory with the effective low energy 
theory at the high scale $\mu$ and evaluating the Wilson coefficients 
from $\mu$ down to the lower scale $\mu\sim O(m_{b})$. In the process
under consideration the high scale $\mu$ is mass of charged Higgs, 
$\mu=m_{H^{\pm}}$. Fortunately, this scale can be taken as the mass of W 
boson, $m_{W}$, since the evaluation from $\mu=m_{H^{\pm}}$ to $\mu=m_{W}$, 
gives negligible contribution to the Wilson coefficients. The reason is that 
the charged Higgs boson is heavy enough from the current theoretical 
restrictions, for example  $m_{H^{\pm}} \geq 340\, GeV$ \cite{ciuchini2}, 
$m_{H^{\pm}} \geq 480\, GeV$ \cite{gudalil}. 

The effective Hamiltonian is obtained by integrating out the heavy degrees 
of freedom, here $t$ quark, $W^{\pm}, H^{\pm}, H_{1}$, and $H_{2}$ bosons 
where $H^{\pm}$ and $H_{1}$,$H_{2}$ denote charged and neutral Higgs 
bosons respectively. For the relevant process we have 
\begin{eqnarray}
{\cal{H}}_{eff}=-4 \frac{G_{F}}{\sqrt{2}} V_{tb} V^{*}_{ts} 
\sum_{i=1}^{12}(C_{i}(\mu) O_{i}(\mu)+C'_{i}(\mu) O'_{i}(\mu)) \, \, ,
\label{hamilton}
\end{eqnarray}
where the $O_{i}$ are current-current ($i=1,2,11,12$), penguin ($i=1,...6$), 
magnetic penguin ($i=7,8$) and semileptonic ($i=9,10$) operators 
\cite{alil3, Grinstein2, misiak} and primed counterparts are their
flipped chirality partners \cite{alil3}.
$C_{i}(\mu)$ and $C'_{i}(\mu)$ are Wilson coefficients renormalized at 
the scale $\mu$. 

Denoting the Wilson coefficients for the SM with $C_{i}^{SM}(m_{W})$ and the
additional charged Higgs contribution with $C_{i}^{H}(m_{W})$, 
we have the initial values for unprimed set of operators 
\cite{alil3} 
\begin{eqnarray}
C^{H}_{1,\dots 6,11,12}(m_W)&=&0 \nonumber \, \, , \\
C_7^{H}(m_W)&=&\frac{1}{m_{t}^2} \,
(\bar{\xi}^{* U}_{N,tt}+\bar{\xi}^{* U}_{N,tc}
\frac{V_{cs}^{*}}{V_{ts}^{*}}) \, (\bar{\xi}^{U}_{N,tt}+\bar{\xi}^{U}_{N,tc}
\frac{V_{cb}}{V_{tb}}) F_{1}(y)\nonumber  \, \, , \\
&+&\frac{1}{m_t m_b} \, (\bar{\xi}^{* U}_{N,tt}+\bar{\xi}^{* U}_{N,tc}
\frac{V_{cs}^{*}}{V_{ts}^{*}}) \, (\bar{\xi}^{D}_{N,bb}+\bar{\xi}^{D}_{N,sb}
\frac{V_{ts}}{V_{tb}}) F_{2}(y)
\nonumber  \, \, , \\
C_8^{H}(m_W)&=&\frac{1}{m_{t}^2} \,
(\bar{\xi}^{* U}_{N,tt}+\bar{\xi}^{* U}_{N,tc}
\frac{V_{cs}^{*}}{V_{ts}^{*}}) \, (\bar{\xi}^{U}_{N,tt}+\bar{\xi}^{U}_{N,tc}
\frac{V_{cb}}{V_{tb}})G_{1}(y)
\nonumber  \, \, , \\
&+&\frac{1}{m_t m_b} \, (\bar{\xi}^{* U}_{N,tt}+\bar{\xi}^{* U}_{N,tc}
\frac{V_{cs}^{*}}{V_{ts}^{*}}) \, (\bar{\xi}^{D}_{N,bb}+\bar{\xi}^{U}_{N,sb}
\frac{V_{ts}}{V_{tb}}) G_{2}(y) \nonumber\, \, , \\
C_9^{H}(m_W)&=&\frac{1}{m_{t}^2} \,
(\bar{\xi}^{* U}_{N,tt}+\bar{\xi}^{* U}_{N,tc}
\frac{V_{cs}^{*}}{V_{ts}^{*}}) \, (\bar{\xi}^{U}_{N,tt}+\bar{\xi}^{U}_{N,tc}
\frac{V_{cb}}{V_{tb}}) H_{1}(y)
\nonumber  \, \, , \\
C_{10}^{H}(m_W)&=&\frac{1}{m_{t}^2} \,
(\bar{\xi}^{* U}_{N,tt}+\bar{\xi}^{* U}_{N,tc}
\frac{V_{cs}^{*}}{V_{ts}^{*}}) \, (\bar{\xi}^{U}_{N,tt}+\bar{\xi}^{U}_{N,tc}
\frac{V_{cb}}{V_{tb}}) L_{1}(y) \, \, , 
\label{CoeffH}
\end{eqnarray}
and for primed set of operators \cite{alil3}, 
\begin{eqnarray}
C^{\prime H}_{1,\dots 6,11,12}(m_W)&=&0 \nonumber \, \, , \\
C^{\prime H}_7(m_W)&=&\frac{1}{m_t^2} \,
(\bar{\xi}^{* D}_{N,bs}\frac{V_{tb}}{V_{ts}^{*}}+\bar{\xi}^{* D}_{N,ss})
\, (\bar{\xi}^{D}_{N,bb}+\bar{\xi}^{D}_{N,sb}
\frac{V_{ts}}{V_{tb}}) F_{1}(y)
\nonumber  \, \, , \\
&+& \frac{1}{m_t m_b}\, (\bar{\xi}^{* D}_{N,bs}\frac{V_{tb}}{V_{ts}^{*}}
+\bar{\xi}^{* D}_{N,ss}) \, (\bar{\xi}^{U}_{N,tt}+\bar{\xi}^{U}_{N,tc}
\frac{V_{cb}}{V_{tb}}) F_{2}(y)
\nonumber  \, \, , \\
C^{\prime H}_8 (m_W)&=&\frac{1}{m_t^2} \,
(\bar{\xi}^{* D}_{N,bs}\frac{V_{tb}}{V_{ts}^{*}}+\bar{\xi}^{* D}_{N,ss})
\, (\bar{\xi}^{D}_{N,bb}+\bar{\xi}^{D}_{N,sb}
\frac{V_{ts}}{V_{tb}}) G_{1}(y)
\nonumber  \, \, , \\
&+&\frac{1}{m_t m_b} \, (\bar{\xi}^{* D}_{N,bs}\frac{V_{tb}}{V_{ts}^{*}}
+\bar{\xi}^{* D}_{N,ss}) \, (\bar{\xi}^{U}_{N,tt}+\bar{\xi}^{U}_{N,tc}
\frac{V_{cb}}{V_{tb}}) G_{2}(y)
\nonumber \,\, ,\\
C^{\prime H}_9(m_W)&=&\frac{1}{m_t^2} \,
(\bar{\xi}^{* D}_{N,bs}\frac{V_{tb}}{V_{ts}^{*}}+\bar{\xi}^{* D}_{N,ss})
\, (\bar{\xi}^{D}_{N,bb}+\bar{\xi}^{D}_{N,sb}
\frac{V_{ts}}{V_{tb}}) H_{1}(y)
\nonumber  \, \, , \\
C^{\prime H}_{10} (m_W)&=&\frac{1}{m_t^2} \,
(\bar{\xi}^{* D}_{N,bs}\frac{V_{tb}}{V_{ts}^{*}}+\bar{\xi}^{D}_{N,ss})
\, (\bar{\xi}^{D}_{N,bb}+\bar{\xi}^{D}_{N,sb}
\frac{V_{ts}}{V_{tb}}) L_{1}(y)
\,\, ,
\label{CoeffH2}
\end{eqnarray}
where $x=m_t^2/m_W^2$ and $y=m_t^2/m_{H^{\pm}}^2$.
In eqs.~(\ref{CoeffH}) and (\ref{CoeffH2}) we used the redefinition
\begin{eqnarray}
\xi^{U,D}=\sqrt{\frac{4 G_{F}}{\sqrt{2}}} \,\, \bar{\xi}^{U,D}\,\, .
\label{ksidefn}
\end{eqnarray}  
The explicit forms of the Wilson coefficients $C_{i}^{(\prime)SM}(m_{W})$
and  the functions $F_{1(2)}(y)$, $G_{1(2)}(y)$, 
$H_{1}(y)$ and $L_{1}(y)$ can be found in Appedix A.
Here we take the couplings $\xi_{ij}^{U,D}$ as complex and neglect the 
contributions due to the neutral Higgs bosons which should be very small
due to the discussion given in \cite{alil2} (see also discussion part).
Finally, the inital values of the Wilson coefficients can be defined as 
\begin {eqnarray}   
C_i^{(')2HDM}(m_{W})&=&C_i^{(')SM}(m_{W})+C_i^{(')H}(m_{W})
\label{CiW}
\end{eqnarray}
Using these initial values, we can calculate the coefficients 
$C_{i}^{2HDM}(\mu)$ and $C^{\prime 2HDM}_{i}(\mu)$ at any lower scale 
in the effective theory with five quarks, namely $u,c,d,s,b$ and use 
the renormalization group to sum the large logaritms, similar to the SM case. 
In this process, Wilson coefficients $C_{7}^{2HDM}(\mu)$, $C_{9}^{2HDM}(\mu)$
and $C_{10}^{2HDM}(\mu)$ play the essential role and the others enter into 
expressions due to operator mixing.

The effective coefficient $C_{7}^{eff}(\mu)$ is defined as 
\begin{eqnarray}
C_{7}^{eff}(\mu)&=&C_{7}^{2HDM}(\mu)+ Q_d \, 
(C_{5}^{2HDM}(\mu) + N_c \, C_{6}^{2HDM}(\mu))\nonumber \, \, , \\
&+& Q_u\, (\frac{m_c}{m_b}\, C_{12}^{2HDM}(\mu) + N_c \, 
\frac{m_c}{m_b}\,C_{11}^{2HDM}(\mu))\nonumber \, \, , \\
C^{\prime eff}_7(\mu)&=& C^{\prime 2HDM}_7(\mu)+Q_{d}\, 
(C^{\prime 2HDM}_5(\mu) + N_c \, C^{\prime 2HDM}_6(\mu))\nonumber \\
&+& Q_u (\frac{m_c}{m_b}\, C_{12}^{\prime 2HDM}(\mu) + N_c \, 
\frac{m_c}{m_b}\,C_{11}^{\prime 2HDM}(\mu))\, \, .
\label{C7eff}
\end{eqnarray}
Here the dependence to coefficients $C_i^{(')2HDM}(\mu)\,\, ,i=5,6,11,12$
comes from the contributions of the operators 
$O_5$, $O_6$, $O_{11}$ and $O_{12}$ ( $O'_5$, $O'_6$, $O'_{11}$ and $O'_{12}$)  
to the leading order matrix element of $b\rightarrow s\gamma$ in the NDR
scheme \cite{alil3}.
The NLO corrected coefficients $C_{7}^{2HDM}(\mu)$ and
$C^{\prime 2HDM}_7(\mu)$  are given as 
\begin{eqnarray}
C_{7}^{2HDM}(\mu)&=&C_{7}^{LO, 2HDM}(\mu) 
+\frac{\alpha_s (\mu)}{4\pi} C_7^{(1)\, 2HDM}(\mu) \nonumber \,\, , \\
C^{\prime 2HDM}_7(\mu)&=& C_{7}^{\prime LO, 2HDM}(\mu) + 
\frac{\alpha_s (\mu)}{4\pi} C_7^{\prime (1)\, 2HDM}(\mu)\,\, .
\label{renwils}
\end{eqnarray}
where the leading order QCD corrected Wilson coefficients
$C_{7}^{LO, 2HDM}(\mu)$ and $C_{7}^{\prime LO, 2HDM}(\mu$)
\cite{buras,Grinstein2,misiak,zakharov}:
\begin{eqnarray} 
C_{7}^{LO, 2HDM}(\mu)&=& \eta^{16/23} C_{7}^{2HDM}(m_{W})+(8/3) 
(\eta^{14/23}-\eta^{16/23}) C_{8}^{2HDM}(m_{W})\nonumber \,\, \\
&+& C_{2}^{2HDM}(m_{W}) \sum_{i=1}^{8} h_{i} \eta^{a_{i}} \,\, , \nonumber
\\
C_{7}^{\prime LO, 2HDM}(\mu)&=& \eta^{16/23} C^{\prime 2HDM}_7(m_{W})+
(8/3) (\eta^{14/23}-\eta^{16/23}) C^{\prime 2HDM}_8(m_{W}) \,\,
\label{LOwils}
\end{eqnarray}
and $\eta =\alpha_{s}(m_{W})/\alpha_{s}(\mu)$, $h_{i}$ and $a_{i}$ are 
the numbers which appear during the evaluation \cite{buras}. 
$C_7^{(1)\, 2HDM}(\mu)$ is the $\alpha_s$ correction to the leading
order result that its explicit form can be found in \cite{ciuchini2,greub2}.
$C_7^{\prime (1)\, 2HDM}(\mu)$ can be obtained by replacing the Wilson 
coefficients in $C_7^{(1)\, 2HDM}(\mu)$ with their primed counterparts. 

The Wilson coefficient $C_9^{eff}(\mu)$ ($C^{\prime eff}_9(\mu)$)
has contributions coming from the coefficients  
$C_1(\mu)$, $C_2(\mu)$, $C_3(\mu)$, ...., $C_6(\mu)$ 
($C^{\prime}_1(\mu)$, $C^{\prime}_2(\mu)$, $C^{\prime}_3
(\mu)$, ..., $C^{\prime}_6(\mu) $) due to the operator mixing.
Therefore the perturbative part of $C_9^{eff}(\mu)$ \cite{buras,misiak} 
and $C^{\prime eff}_9(\mu)$ including NLO QCD corrections are defined 
in the NDR scheme as:  
\begin{eqnarray} 
C_9^{pert}(\mu)&=& C_9^{2HDM}(\mu) \tilde\eta (\hat s) \nonumber 
\\ &+& h(z, \hat s) \left( 3 C_1(\mu) + C_2(\mu) + 3 C_3(\mu) + 
C_4(\mu) + 3 C_5(\mu) + C_6(\mu) \right) \nonumber \\
&- & \frac{1}{2} h(1, \hat s) \left( 4 C_3(\mu) + 4 C_4(\mu) + 3
C_5(\mu) + C_6(\mu) \right) \\
&- &  \frac{1}{2} h(0, \hat s) \left( C_3(\mu) + 3 C_4(\mu) \right) +
\frac{2}{9} \left( 3 C_3(\mu) + C_4(\mu) + 3 C_5(\mu) + C_6(\mu)
\right) \nonumber \,\, ,
\label{C9eff}
\end{eqnarray}
and
\begin{eqnarray} 
C_9^{\prime\, pert}(\mu)&=& C_9^{\prime 2HDM}(\mu) \tilde\eta(\hat s)\nonumber 
\\ &+& h(z, \hat s) \left( 3 C'_1(\mu) + C'_2(\mu) + 3 C'_3(\mu) + 
C'_4(\mu) + 3 C'_5(\mu) + C'_6(\mu) \right) \nonumber \\
&-& \frac{1}{2} h(1, \hat s) \left( 4 C'_3(\mu) + 4 C'_4(\mu) + 3
C'_5(\mu) + C'_6(\mu) \right) \\
&-&  \frac{1}{2} h(0, \hat s) \left( C'_3(\mu) + 3 C'_4(\mu) \right) +
\frac{2}{9} \left( 3 C'_3(\mu) + C'_4(\mu) + 3 C'_5(\mu) + C'_6(\mu)
\right) \nonumber \, .
\label{C9effp} 
\end{eqnarray}
where $z=\frac{m_c}{m_b}$ and $\hat s=\frac{q^2}{m_b^2}$.
In the above expression $\tilde\eta(\hat s)$ represents the one gluon
correction to the matrix element $O_9$ with $m_s=0$ \cite{misiak} and
the function $h(z,\hat s)$ arises from the one loop contributions of the
four quark operators $O_1, ... ,O_6$ ($O'_1, ... ,O'_6$) (see Appendix B).
There exist also the long distance (LD) part due to the real $\bar{c}c$ in 
the intermediate states, i.e. the cascade process 
$B\rightarrow K^* \psi_i \rightarrow K^* l^+ l^-$ where $i=1,..,6$. 
Using a Breit-Wigner form of the resonance propogator \cite{donnel,zakharov},
and adding this contribution to the perturbative one coming from the 
$c\bar{c}$ loop, the NLO QCD corrected $C_9^{eff}(\mu)$ can be
written as: 
\begin{eqnarray}
C_9^{eff}(\mu)=C_9^{pert}(\mu)+ Y_{reson}(\hat{s})\,\, ,
\label{C9efftot}
\end{eqnarray}
where $Y_{reson}(\hat{s})$ in NDR scheme is defined as
\begin{eqnarray}
Y_{reson}(\hat{s})&=&-\frac{3}{\alpha^2_{em}}\kappa \sum_{V_i=\psi_i}
\frac{\pi \Gamma(V_i\rightarrow ll)m_{V_i}}{q^2-m_{V_i}+i m_{V_i}
\Gamma_{V_i}} \nonumber \\
& & \left( 3 C_1(\mu) + C_2(\mu) + 3 C_3(\mu) + 
C_4(\mu) + 3 C_5(\mu) + C_6(\mu) \right).
\label{Yres}
\end{eqnarray}
For the expression $C_9^{\prime \, eff}(\mu)$, it is enough to replace
all unprimed coefficients with primed ones. In eq. (\ref{Yres}) the 
phenomenological parameter $\kappa=2.3$ is chosen \cite{ali}.
The NLO corrected coefficients $C_i\,\,, i=1,...,6$ can be found in 
\cite{ciuchini2,greub2}. 

Finally, neglecting the strange quark mass, the matrix element for 
$b \rightarrow s \ell^+\ell^-$ decay is obtained as:
\begin{eqnarray}
{\cal M}&=& - \frac{G_F \alpha_{em}}{2\sqrt 2 \pi} V_{tb} V^*_{ts} 
\Bigg\{ \left( \, C_9^{eff}(\mu)\,
\bar s \gamma_\mu (1- \gamma_5) b + 
C_9^{\prime eff}(\mu)\, \bar s \gamma_\mu (1+ \gamma_5) b \, \right)
\,\, \bar \ell \gamma^\mu \ell \nonumber \\
&+& \left( \, C_{10}(\mu) \, \bar s \gamma_\mu (1- \gamma_5) b+
C'_{10}(\mu)\, \bar s \gamma_\mu (1+ \gamma_5) b \, \right) \,\,
\bar \ell \gamma^\mu \gamma_5 \ell   \\
&-& 2 \left( \, C^{eff}_7(\mu)\, \frac{m_b}{q^2}\, 
\bar s i \sigma_{\mu \nu}q^\nu (1+\gamma_5)  b
+C^{\prime eff}_7(\mu)\, \frac{m_b}{q^2}\, 
\bar s i \sigma_{\mu \nu}q^\nu (1-\gamma_5)
b \, \right) \,\, \bar \ell \gamma^\mu \ell \Bigg\}~\nonumber .
\label{matr}
\end{eqnarray}

To look at the problem from the hadronic side, the 
$B\rightarrow K^* l^+ l^-$ decay, it is necessary to calculate the matrix 
elements
$ \la K^* \vel \bar s \gamma_\mu (1\pm \gamma_5) b \ver B \ra$, and
$\la K^* \vel \bar s i \sigma_{\mu \nu} q^\nu (1\pm\gamma_5) b \ver B \ra$.
Using the parametrization of the form factors as in \cite{colangelo}, 
the matrix element of the $B\rightarrow K^* l^+ l^-$ decay is obtained as 
\cite{alsav2}:
\begin{eqnarray}
{\cal M} &=& -\frac{G \alpha_{em}}{2 \sqrt 2 \pi} V_{tb} V_{ts}^*  
\Bigg\{ \bar \ell \gamma^\mu
\ell \left[ 2 A_{tot} \epsilon_{\mu \nu \rho \sigma} \epsilon^{* \nu} 
p_{K^*}^\rho q^\sigma + i
B_{1 \,tot} \epsilon^*_\mu - i B_{2 \,tot} ( \epsilon^* q) 
(p_{B}+p_{K^*})_\mu - 
i B_{3\, tot} (\epsilon^* q)q_\mu \right] \nonumber \\
&+& \bar \ell \gamma^\mu \gamma_5 \ell \left[ 2 C_{tot} \epsilon_{\mu \nu \rho
\sigma}\epsilon^{* \nu} p_{K^*}^\rho q^\sigma + i D_{1\, tot} \epsilon^*_\mu - 
i D_{2\, tot} (\epsilon^* q) (p_{B}+p_{K^*})_\mu - i D_{3\, tot} 
(\epsilon^* q) q_\mu \right] \Bigg\}~,
\label{matr2}
\end{eqnarray}
where $\epsilon^{* \mu}$ is the polarization vector of $K^*$ meson, $p_{B}$ 
and $p_{K^*}$ are four momentum vectors of $B$ and $K^*$ mesons, 
$q=p_B-p_{K^*}$. $A_{tot}$, $C_{tot}$, $B_{i\, tot}$, and  $D_{i\, tot}$ 
$i=1,2,3$  are functions of Wilson coefficients and form factors of the
relevant process. Their explicit forms are given in Appendix C. 

Now we are ready to calculate the CP-violating asymmetry for the given 
process. The complex Yukawa couplings are the possible source of CP 
violation in the model III for the decay $B\rightarrow K^* l^+ l^-$.
In our theoretical calculations we expect that the neutral Higgs effects
on the Wilson coefficent $C_7^{eff}$ is suppressed and we neglect all the 
Yukawa couplings, except $\bar{\xi}^{U}_{N,tt}$ and $\bar{\xi}^{D}_{N,bb}$
(see discussion part). Therefore, in model III, the only detectable CP 
violating effect comes from the combination 
$\bar{\xi}^{U *}_{N,tt} \bar{\xi}^{D}_{N,bb}$, appears in the Wilson 
coefficient $C_7^{eff}$. 
Using the definition of CP-violating asymmetry ($A_{CP}$)
\begin{eqnarray}
A_{CP}= \frac{\frac{d \Gamma (\bar{B_s}\rightarrow K^* e^+ e^-)}{dq^2 }-
\frac{d \Gamma (B_s\rightarrow \bar{K^*} e^+ e^-)}{dq^2 }}
{\frac{d \Gamma (\bar{B_s}\rightarrow K^* e^+ e^-)}{dq^2 }+
\frac{d \Gamma (B_s\rightarrow \bar{K^*} e^+ e^-)}{dq^2 }}\,\, .
\label{cpvio}
\end{eqnarray}
we get 
\begin{eqnarray}
A_{CP}=-2 Im (\lambda_2) \frac{Im (C_9^{eff}(m_b)) \,\, P_1\, \, \Delta}
{Re(\lambda_2)[-2 (P_1+2 P_2)\, Re (C_9^{eff}(m_b))\,\Delta +\Omega}.
\label{Acp}
\end{eqnarray}
In eq. (\ref{Acp}) we use the parametrization
\begin{eqnarray}
C_7^{eff}(\mu)=P_1(\mu) \, \lambda_2 + P_2(\mu)\,\,,
\label{param}
\end{eqnarray}
where $\lambda_2$ is
\begin{eqnarray}
\lambda_2=\frac{1}{m_t\, m_b}|\bar{\xi}^{U}_{N,tt} \bar{\xi}^{D}_{N,bb}|
e^{i\theta}
\label{lambda2}
\end{eqnarray}
Note that, here, we choose $\bar{\xi}^{U}_{N,tt}$ as real and 
$\bar{\xi}^{D}_{N,bb}$ as complex, namely 
$\bar{\xi}^{D}_{N,bb}=|\bar{\xi}^{D}_{N,bb}|\, e^{i \theta}$ 
(see discussion).   
Functions $P_1(\mu)$ and $P_2(\mu)$ can be written as the combinations of 
LO and NLO part, namely,
\begin{eqnarray}
P_1(\mu)&=&P_1^{LO}(\mu)+P_1^{NLO}(\mu)\nonumber \,\, ,\\
P_2(\mu)&=&P_2^{LO}(\mu)+P_2^{NLO}(\mu)\,\, , 
\label{P12}
\end{eqnarray}
and 
\begin{eqnarray}
P_1^{LO}(\mu)&=&\eta^{16/23}
F_2(y)+\frac{8}{3}(\eta^{14/23}-\eta^{16/23})\,G_2(y)\nonumber \\
P_2^{LO}(\mu)&=&\eta^{16/23}
[C_7^{SM}(m_W)+\frac{|\bar{\xi}^{U}_{N,tt}|^2}{m_t^2}  F_1(y)]\nonumber \\
&+&\frac{8}{3}(\eta^{14/23}-\eta^{16/23})
[C_8^{SM}(m_W)+\frac{|\bar{\xi}^{U}_{N,tt}|^2}{m_t^2}  G_1(y)]\nonumber \\
&+&Q_d (C_5^{LO}(\mu)+N_c \, C_6^{LO}(\mu))+
Q_u (\frac{m_c}{m_b} C_{12}^{LO}(\mu)+N_c \frac{m_c}{m_b} C_{11}^{LO}(\mu))
\nonumber\\
&+& C_2(m_W) \sum_{i=1}^{8} h_i \eta^{a_i}
\label{P12LO}
\end{eqnarray}
$P_1^{NLO}(\mu)$ is calculated by extracting the coefficient of 
$\lambda_2$ in the expression 
$\frac{\alpha_s(\mu)}{4 \pi} C_7^{(1)\, 2HDM)}(\mu)$. 
Similarly, $P_2^{NLO}(\mu)$ is obtained by setting $\lambda_2=0$ in the 
same expression. Finally, the functions $\Delta$ and $\Omega$ are defined 
as
\begin{eqnarray}
\Delta&=& -\frac{T_2 s}{3 q^2 r (1+\sqrt r)}\Bigg \{ A_2 \lambda (-1-3r+s)+
A_1 (1+\sqrt r)^2(\lambda-12 r (r-1))\Bigg \} \nonumber \\ 
&+& \frac{T_3 \lambda}{3 m_B^2 r (1+\sqrt r)(r-1)}\Bigg 
\{ A_2 \lambda + A_1 (1+\sqrt r)^2 (-1+r+s))\Bigg \}
-\frac{8 T_1 V s}{3 q^2 (1+\sqrt r)}\lambda  \nonumber \,\, ,\\
\Omega&=& \frac{|C_9^{eff}|^2+|C_{10}|^2}{6 m_b m_B (1+\sqrt r)^2 r}
\Bigg \{ 2 A_1 A_2 \lambda (1+\sqrt r)^2 (-1+r+s)+
A_1^2 (1+\sqrt r)^4 (\lambda+12 r s)\nonumber \\ 
&+& \lambda^2 A_2^2+8 \lambda r s V^2 \Bigg \} \nonumber \\ 
&+& 8 (P_1+P_2) P_2 \Bigg \{ \frac{8 \lambda m_b m_B}{3 q^4} T_1^2 s 
+ \frac{m_b m_B}{3 q^4 r} T_2^2 [\lambda (-4 r+s)+ 12 r (r-1)^2]\, s
\nonumber \\ 
&+& \frac{m_b }{3 m_B^3 r (-1+r)^2} \lambda^2 T_3^2 \nonumber \\
&+& \frac{2 \lambda m_b }{3 m_B q^2r (-1+r)} s (1-s+3 r)T_2 T_3 \Bigg \}
\label{DelOme}
\end{eqnarray}
where $r=\frac{m^2_{K^*}}{m^2_{B}}$, $s=\frac{q^2}{m^2_B}$ and 
$\lambda=1+r^2+s^2-2 r-2 s- 2 r s$.
\section{Discussion}
Model III induces many free parameters, namely, complex Yukawa couplings, 
$\xi_{ij}^{U,D}$ where i,j are flavor indices, masses of charged and neutral 
Higgs bosons. These parameters should be restricted using the 
experimental measurements. The contributions of the neutral Higgs bosons 
$h_0$ and $A_0$ to the Wilson coefficient $C_7^{eff}$ (see the appendix of 
\cite{alil2} for details) are not in contradiction with the CLEO measurement
announced recently  
\cite{cleo2}, 
\begin{eqnarray}
Br (B\rightarrow X_s\gamma)= (3.15\pm 0.35\pm 0.32)\, 10^{-4} \,\, ,
\label{br2}
\end{eqnarray}
if the couplings $\bar{\xi}^{D}_{N,is}$($i=d,s,b)$ and 
$\bar{\xi}^{D}_{N,db}$ are negligible.
Further, using the constraints \cite{alil1}, 
coming from the $\Delta F=2$ mixing, the $\rho$ parameter \cite{atwood}, 
and the measurement by CLEO Collaboration, we have :   
$\bar{\xi}_{N tc} << \bar{\xi}^{U}_{N tt},
\,\,\bar{\xi}^{D}_{N bb}$ and $\bar{\xi}^{D}_{N ib} \sim 0\, , 
\bar{\xi}^{D}_{N ij}\sim 0$, where the indices $i,j$ denote d and s quarks .    
These restrictions allows us to neglect all the couplings except 
$\bar{\xi}^{U}_{N tt}$ and $\bar{\xi}^{D}_{N bb}$.
With this choice, we can cancel the contributions coming from 
primed Wilson coefficients eq.(\ref{CoeffH2}) and the neutral Higgs  bosons
since the Yukawa vertices are combinations of 
$\bar{\xi}^{D}_{N sb}$ and $\bar{\xi}^{D}_{N ss}$.
Finally, only the multiplication of Yukawa couplings, 
$\bar{\xi}^{U}_{N tt}\,\bar{\xi}^{* D}_{N bb}$ and  
$|\bar{\xi}^{U}_{N tt}|^2$ appear in the Wilson coefficients 
(see eq. \ref{CoeffH}). At this stage it is possible to define a new 
parameter $\theta$ with the expression 
\begin{eqnarray}   
\bar{\xi}^{U}_{N tt}\,\bar{\xi}^{* D}_{N bb}=
|\bar{\xi}^{U}_{N tt}\,\bar{\xi}^{* D}_{N bb}| e^{-i\theta}
\label{theta}
\end{eqnarray}

Here, it is possible to take both $\bar{\xi}^{U}_{N,tt}$ and  
$\bar{\xi}^{D}_{N,bb}$ or any one of them complex. In our work, we choose 
$\bar{\xi}^{U}_{N,tt}$ as real and  $\bar{\xi}^{D}_{N,bb}$ as complex,
namely $\bar{\xi}^{D}_{N,bb}=|\bar{\xi}^{D}_{N,bb}|\, e^{i \theta}$. 

The phase angle $\theta$ leads to a substantial enhancement in
neutron electric dipole moment and 
the experimental upper limit on neutron electric dipole moment 
$d_n<10^{-25}\hbox{e$\cdot$cm}$ thus places
a upper bound on the couplings: 
$\frac{1}{m_t m_b} Im(\bar{\xi}^{U}_{N tt}\,\bar{\xi}^{* D}_{N bb})< 1.0$ 
for $M_{H^\pm}\approx 200$ GeV \cite{david}.

In this section, we study the $q^2$ dependencies of the CP asymmetry 
$A_{CP}$ of the decay $B\rightarrow K^* l^+ l^-$ 
for the selected parameters of the model III
($\bar{\xi}^{U}_{N tt}$,  $\bar{\xi}^{D}_{N bb}$ and phase angle $\theta$)
In our analysis, we restricted $|C_7^{eff}|$ in the region 
$0.257 \leq |C_7^{eff}| \leq 0.439$ coming from CLEO measurement 
\cite{cleo2}. Here upper and lower limits were calculated in \cite{alil1}
following the procedure given in \cite{gudalil}. With this restriction,
an allowed region for the parameters  $\bar{\xi}^{U}_{N tt}$, 
$\bar{\xi}^{D}_{N bb}$ and $\theta$, is found. 
Our numerical calculations based on this restriction and the constraint for
the angle $\theta$ due to the experimental upper limit of neutron electric
dipole moment. Throughout these calculations, we take the charged Higgs mass
$m_{H^{\pm}}=400\, GeV$, the scale $\mu=m_b$ and we use the input values
given in Table (\ref{input}).  
\begin{table}[h]
        \begin{center}
        \begin{tabular}{|l|l|}
        \hline
        \multicolumn{1}{|c|}{Parameter} & 
                \multicolumn{1}{|c|}{Value}     \\
        \hline \hline
        $m_c$                   & $1.4$ (GeV) \\
        $m_b$                   & $4.8$ (GeV) \\
        $\alpha_{em}^{-1}$      & 129           \\
        $\lambda_t$            & 0.04 \\
        $m_{B_d}$             & $5.28$ (GeV) \\
        $m_{t}$             & $175$ (GeV) \\
        $m_{W}$             & $80.26$ (GeV) \\
        $m_{Z}$             & $91.19$ (GeV) \\
        $\Lambda_{QCD}$             & $0.214$ (GeV) \\
        $\alpha_{s}(m_Z)$             & $0.117$  \\
        $sin\theta_W$             & $0.2325$  \\
        \hline
        \end{tabular}
        \end{center}
\caption{The values of the input parameters used in the numerical
          calculations.}
\label{input}
\end{table}

\begin{figure}[htb]
\vskip -3.0truein
\centering
\epsfxsize=6.8in
\leavevmode\epsffile{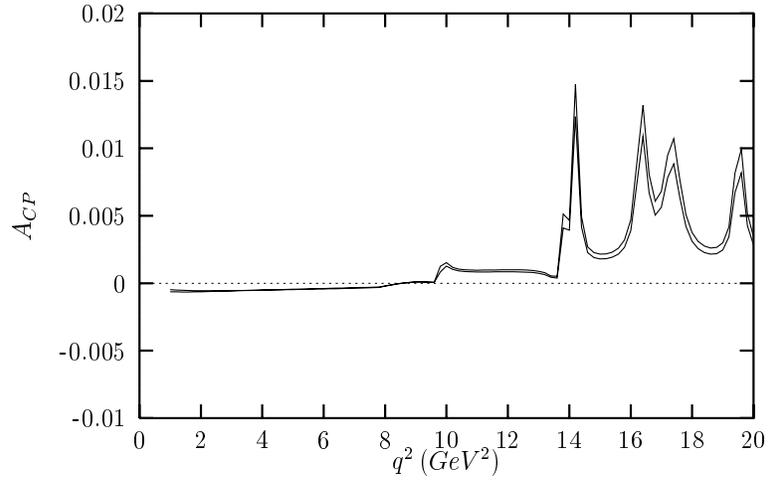}
\vskip -3.0truein
\caption[]{$A_{CP}$ as a function of  $q^2$ 
for fixed $\bar{\xi}_{N,bb}^{D}=40\, m_b$ in the region $|r_{tb}|<1$,
at the scale $\mu=m_b$, for $C_7^{eff}>0$ and $sin\theta =0.1$, including 
LD effects. Here $A_{CP}$ is restricted in the region bounded by solid 
lines .} 
\label{ACPIII4001pl}
\end{figure}

\begin{figure}[htb]
\vskip -3.0truein
\centering
\epsfxsize=6.8in
\leavevmode\epsffile{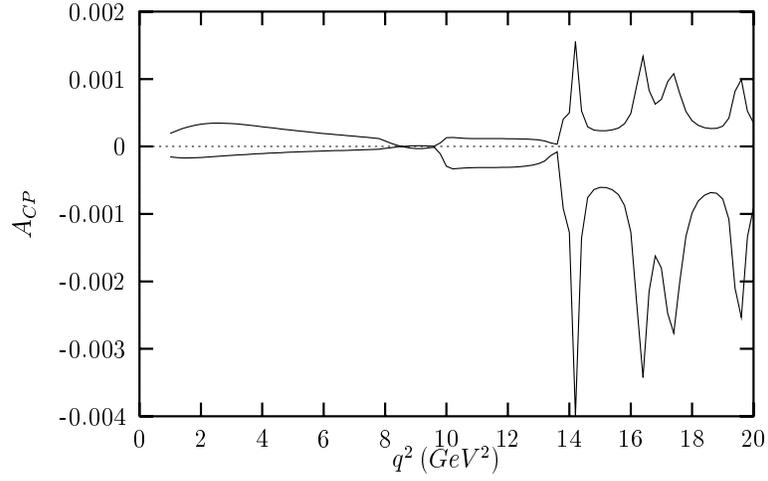}
\vskip -3.0truein
\caption[]{The same as Fig 1, but for $C_7^{eff} < 0$ .}
\label{ACPIII4001mn}
\end{figure}

\begin{figure}[htb]
\vskip -3.0truein
\centering
\epsfxsize=6.8in
\leavevmode\epsffile{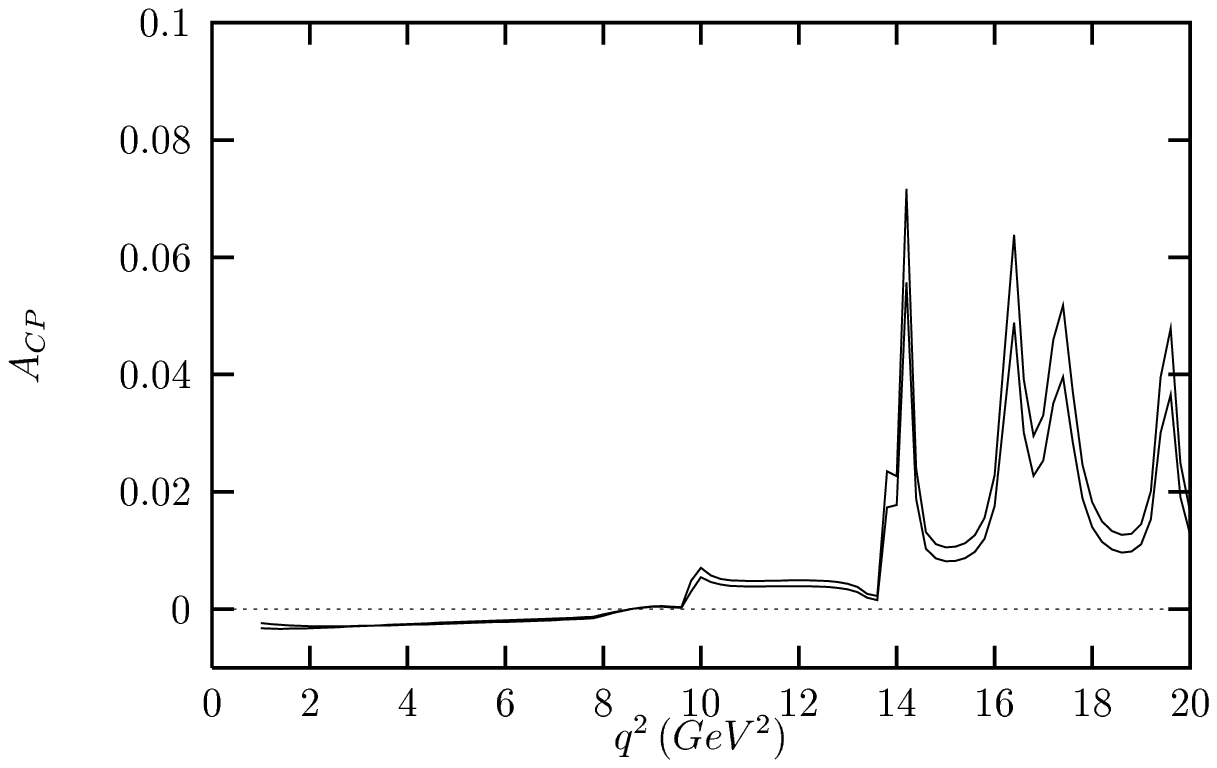}
\vskip -3.0truein
\caption[]{The same as Fig 1, 
but for $sin\,\theta =0.5$.}
\label{ACPIII4005pl}
\end{figure}

\begin{figure}[htb]
\vskip -3.0truein
\centering
\epsfxsize=6.8in
\leavevmode\epsffile{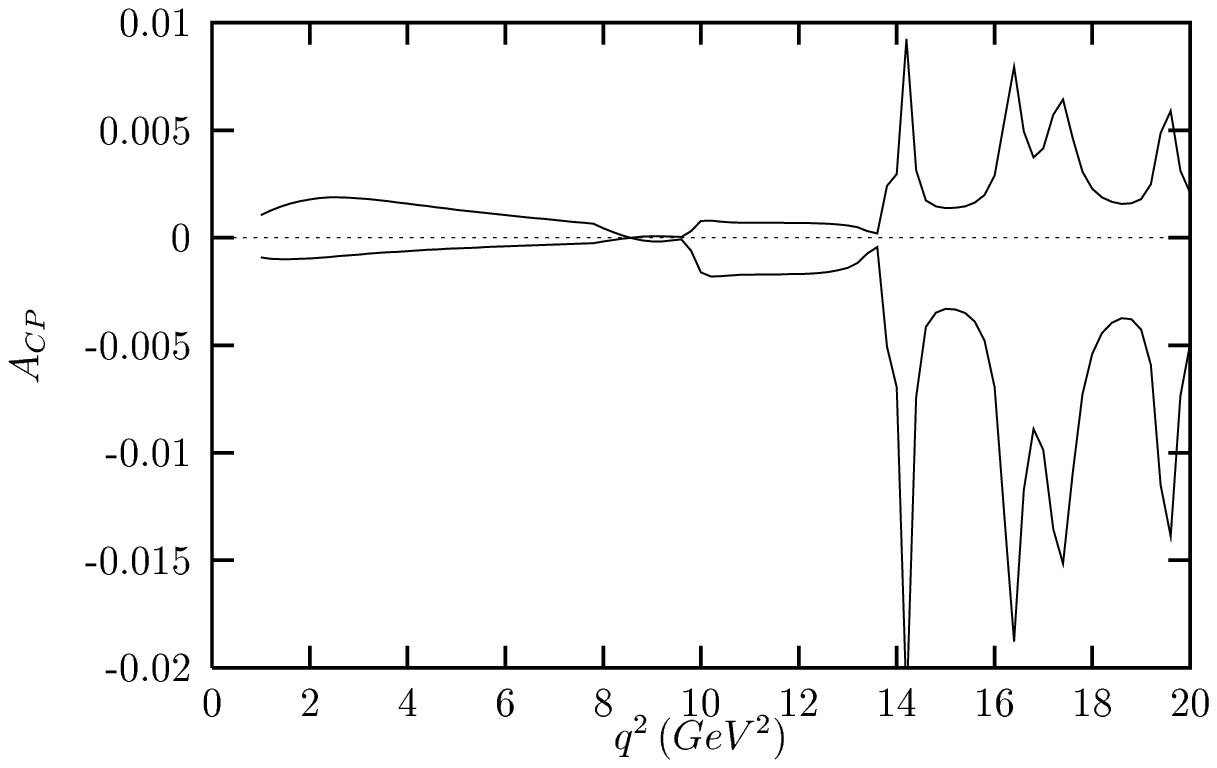}
\vskip -3.0truein
\caption[]{The same as Fig 2, but for $sin\,\theta =0.5$.}
\label{ACPIII4005mn}
\end{figure}

\begin{figure}[htb]
\vskip -3.0truein
\centering
\epsfxsize=6.8in
\leavevmode\epsffile{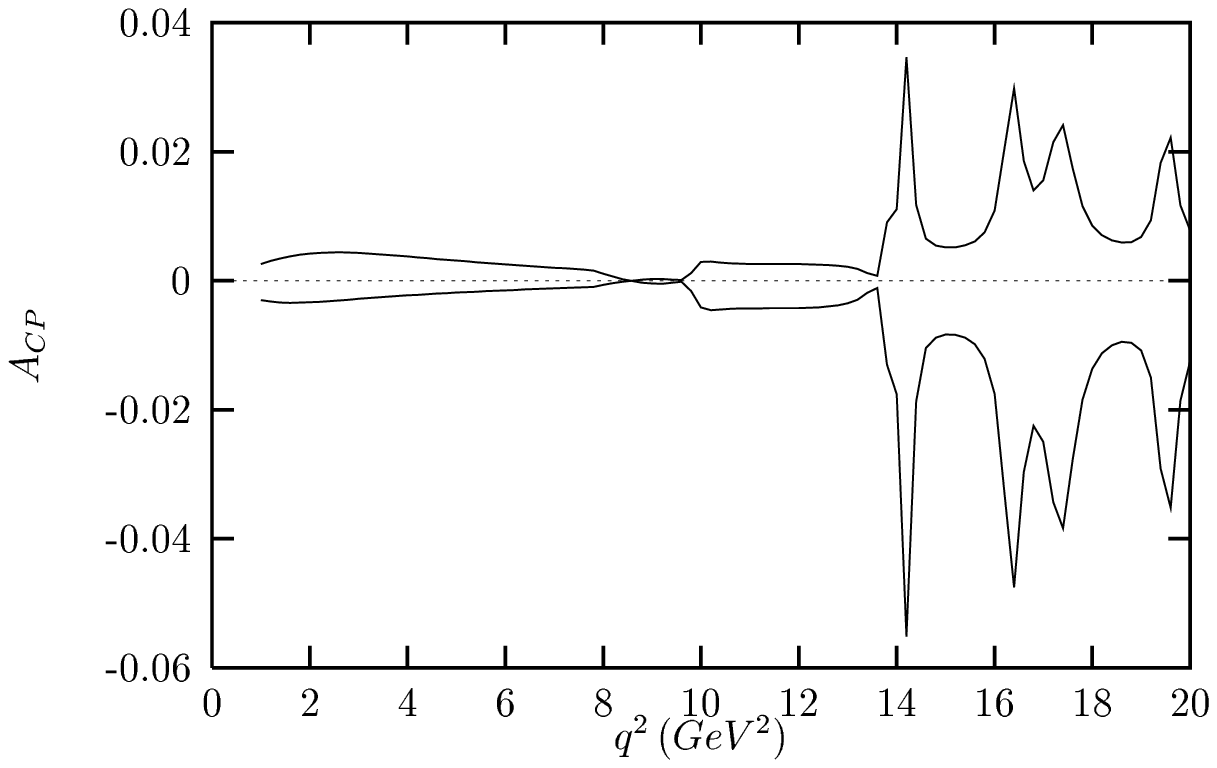}
\vskip -3.0truein
\caption[]{The same as Fig 2, but for $sin\,\theta =0.9$.}
\label{ACPIII4009mn}
\end{figure}

\begin{figure}[htb]
\vskip -3.0truein
\centering
\epsfxsize=6.8in
\leavevmode\epsffile{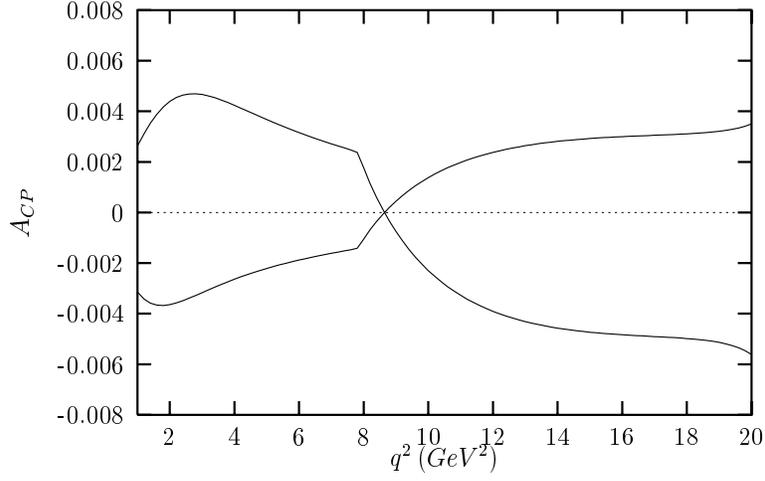}
\vskip -3.0truein
\caption[]{The same as Fig \ref{ACPIII4009mn}, but withouth LD effects.}
\label{ACPIII4009mn0}
\end{figure}

In  fig.~\ref{ACPIII4001pl} we plot $A_{CP}$ 
of the decay $B\rightarrow K^* l^+ l^-$ with respect to the dilepton
mass square, $q^2$, for $\bar{\xi}_{N,bb}^{D}=40\, m_b$, $sin\,\theta=0.1$ 
and $C_7^{eff} > 0$, in the case where the ratio 
$|r_{tb}|=|\frac{\bar{\xi}_{N,tt}^{U}}{\bar{\xi}_{N,bb}^{D}}| <1.$
$A_{CP}$ is restricted in the narrow region bounded by solid lines, 
which almost coincide, especially for the values of $q^2$ far from 
resonances. Up to the value $q^2=8\, GeV^2$ $A_{CP}$ is negative, 
however it changes sign almost at $q^2=9\, GeV^2$.
In  fig.~\ref{ACPIII4001mn}, we present the same dependence as in 
fig.~\ref{ACPIII4001pl} for $C_7^{eff} < 0$. Here, $A_{CP}$ lies in the
region bounded by solid lines and it can change sign for any $q^2$ value.
In both cases, $C_7^{eff} > 0$ and $C_7^{eff} < 0$, $A_{CP}$ is small, at
order of $10^{-3}$ in the region far from resonances. 
Fig. \ref{ACPIII4005pl} (\ref{ACPIII4005mn}) show the same dependence like 
fig. \ref{ACPIII4001pl} (\ref{ACPIII4001mn}), but for $sin\,\theta =0.5$.
$A_{CP}$ behaves similar to $sin\,\theta$ case, however it increases 
almost $5$ times compared to former one. Further, the restriction region for
$A_{CP}$ becomes broader. We also show the $q^2$ dependence of $A_{CP}$ 
for $sin\theta =0.9$  (withouth LD effects) in fig. 
\ref{ACPIII4009mn}(\ref{ACPIII4009mn0}) for $C_7^{eff} < 0$. 
It should be noted that, for $C_7^{eff} > 0$, $sin\,\theta=0.9$ does not obey 
the restriction coming from the limit on neutron electric dipole moment, 
namely 
$\frac{1}{m_t m_b} Im(\bar{\xi}^{U}_{N tt}\,\bar{\xi}^{* D}_{N bb})< 1.0$ .
However, this restriction is still satisfied for $C_7^{eff} < 0$  
(fig. \ref{ACPIII4009mn} and \ref{ACPIII4009mn0}), since the ratio 
$|\frac{\bar{\xi}^{U}_{N tt}}{\bar{\xi}^{* D}_{N bb}}|$ is small enough. 
Finally,  for the case $|r_{tb}|>>1$, $sin\,\theta$ should be very small
to satisfy the neutron electric dipole moment restriction and $A_{CP}$ 
almost vanishes ($\sim 10^{-11}$).
Now, we present the average $A_{CP}$,$\bar{A}_{CP}$, for three different 
phase angles ($sin\theta=0.1,\,0,5,\,0.9$) in two different dilepton mass 
regions (Table \ref{Table1}).
\begin{table}[h]
\small{    \begin{center}
    \begin{tabular}{|c|c|c|c|c|}
    \hline
    \hline \hline
    $sin\theta$   &$C_7^{eff}>0$&  $C_7^{eff}<0$ &$q^2$ regions\\
\hline \hline
    $0.1$            &$-3.32\, 10^{-3}\leq \bar{A}_{CP}\leq -3.35\, 10^{-3} $ 
&$-0.70\, 10^{-3}\leq \bar{A}_{CP}\leq 1.72\, 10^{-3} $& I  \\
    \hline
&$3.16\, 10^{-3}\leq \bar{A}_{CP}\leq 3.75\, 10^{-3} $ 
&$-1.06\, 10^{-3}\leq \bar{A}_{CP}\leq 0.41\, 10^{-3} $ & II \\
    \hline \hline

    $0.5$            &$-1.68\, 10^{-2}\leq \bar{A}_{CP}\leq -1.66\, 10^{-2} $ 
&$-4.18\, 10^{-3}\leq \bar{A}_{CP}\leq 9.31\, 10^{-3} $& I  \\
    \hline
&$1.43\, 10^{-2}\leq \bar{A}_{CP}\leq 1.82\, 10^{-2} $ 
&$-5.80\, 10^{-3}\leq \bar{A}_{CP}\leq 2.42\, 10^{-3} $ & II \\
    \hline \hline
    $0.9$            & 
&$-1.49\, 10^{-2}\leq \bar{A}_{CP}\leq 2.22\, 10^{-2} $& I  \\
    \hline
& 
&$-1.46\, 10^{-2}\leq \bar{A}_{CP}\leq 9.07\, 10^{-3} $ & II \\
    \hline \hline
        \end{tabular}
        \end{center} }
\caption{ The average asymmetry $\bar{A}_{CP}$ for regions I 
( $1\, GeV\leq \sqrt q^2 \leq m_{J/\psi}-20\, MeV$ ) and II 
($m_{J/\psi}+20\, MeV \leq \sqrt q^2 \leq m_{\psi'}-20\, MeV$ ) }
\label{Table1}
\end{table}

\begin{figure}[htb]
\vskip -3.0truein
\centering
\epsfxsize=6.8in
\leavevmode\epsffile{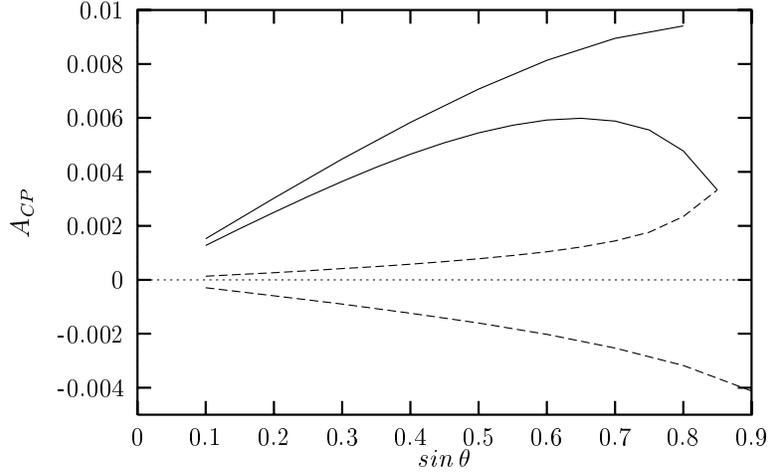}
\vskip -3.0truein
\caption[]{$A_{CP}$ as a function of  $sin\,\theta$  for $q^2=10\, GeV^2$, 
$\bar{\xi}_{N,bb}^{D}=40\, m_b$ in the region $|r_{tb}|<1$,
at the scale $\mu=m_b$. For $C_7^{eff} > 0$, $A_{CP}$ lies in the region 
bounded by solid lines and for $C_7^{eff}<0$, it lies in the region bounded 
by dashed lines .}
\label{ACPIII40q210}
\end{figure}

\begin{figure}[htb]
\vskip -3.0truein
\centering
\epsfxsize=6.8in
\leavevmode\epsffile{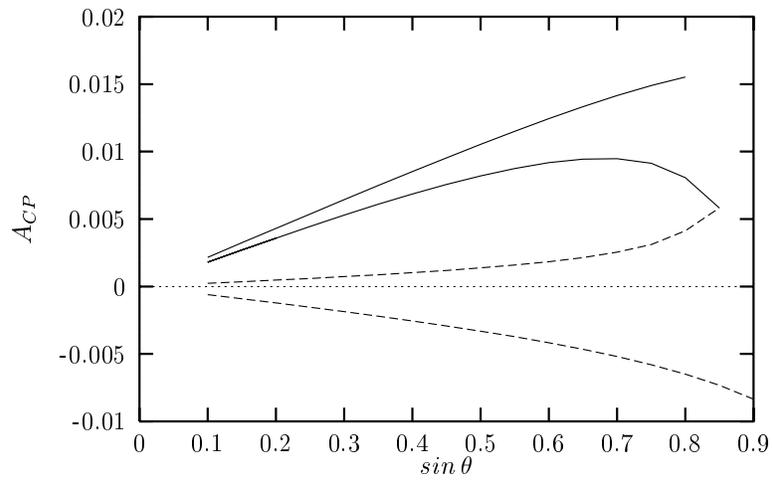}
\vskip -3.0truein
\caption[]{The same as Fig \ref{ACPIII40q210}, but for $q^2=15\,GeV^2$.}
\label{ACPIII40q215}

\end{figure}

Figs. \ref{ACPIII40q210} and \ref{ACPIII40q215} are devoted to $sin\,\theta$
dependence of $A_{CP}$ with LD effects for $q^2=10\, GeV^2$ and 
$q^2=15\, GeV^2$ respectively. Here, $A_{CP}$ lies in the region bounded 
by solid lines for $C_7^{eff} > 0$ or by dashed lines for $C_7^{eff} < 0$.
It is interesting that the lower bound of the region for $C_7^{eff} > 0$
coincides with the upper bound of the region for $C_7^{eff} < 0$, at almost
$sin\,\theta=0.8$. Decreasing $sin\,\theta$ causes to decrease $A_{CP}$ as
expected and it makes the restricted region narrower,
for both $C_7^{eff} < 0$ and $C_7^{eff} < 0$.
Further, for $q^2=10\, GeV^2$ and $q^2=15\, GeV^2$, $A_{CP}$ is positive 
for $C_7^{eff} > 0$. However it can have negative values for $C_7^{eff} < 0$. 
This is informative in the determination of the sign of $C_7^{eff}$.

In conclusion, we analyse the dependency of, $A_{CP}$ on $q^2$ and 
$sin\,\theta$ using the restrictions for the model III parameters 
$\bar{\xi}_{N,tt}^{U}\,, \bar{\xi}_{N,bb}^{D}$, $sin\,\theta$ and 
we calculate the average CP-asymmetry $\bar{A}_{CP}$ 
in two different dilepton mass regions, for the decay 
$B\rightarrow K^* l^+ l^-$. 

Now we would like to summarize the main points of our results:

\begin{itemize}
\item For $|r_{tb}| <1$ and $C_{7}^{eff}>0$, increasing $sin\, \theta$ 
causes to increase $|A_{CP}|$ and the area of the restriction region. 
In this case, $A_{CP}$ changes sign at the $q^2$ value, 
$q^{2}\sim 9 \, GeV^2$. For $|r_{tb}| <1$ and $C_{7}^{eff}<0$ the 
restriction region becomes broader with increasing $sin\,\theta$, however 
$A_{CP}$ can be very small and even vanish for any value of $q^2$.

\item  For the case $|r_{tb}|>>1$, $A_{CP}$ almost 
vanishes ($\sim 10^{-11}$) since $sin\,\theta$ should be very small
due to the restriction coming from the limit on neutron electric dipole 
moment. 

Therefore, if $A_{CP}$ is not observed for the relevant process, it is still 
possible to have physics beyond the SM, here the general 2HDM with    
$|r_{tb}| <1$ , $C_{7}^{eff} < 0$ or $|r_{tb}| >>1$.  

\item  For the fixed value of $q^2$, $q^2=10\,GeV^2$ (or $q^2=15\,GeV^2$),
$C_{7}^{eff}$ can have both signs if $A_{CP}$ is measured as positive. 
However, if $A_{CP}$ is negative, $C_{7}^{eff}$ will be negative. 
This shows that the measurement of $A_{CP}$ for fixed $q^{2}$  gives 
information about the sign of $C_{7}^{eff}$, which is an interesting result.

\end{itemize}

Therefore, the experimental investigation of $A_{CP}$ ensure a crucial test
for new physics and also the sign of $C_{7}^{eff}$.
\newpage
{\bf \LARGE {Appendix}} \\
\begin{appendix}
\section{The Wilson coefficients in the SM and the functions appear in
these coefficients}
The initial values of the Wilson coefficients for the relevant process 
in the SM are \cite{Grinstein1}
\begin{eqnarray}
C^{SM}_{1,3,\dots 6,11,12}(m_W)&=&0 \nonumber \, \, , \\
C^{SM}_2(m_W)&=&1 \nonumber \, \, , \\
C_7^{SM}(m_W)&=&\frac{3 x^3-2 x^2}{4(x-1)^4} \ln x+
\frac{-8x^3-5 x^2+7 x}{24 (x-1)^3} \nonumber \, \, , \\
C_8^{SM}(m_W)&=&-\frac{3 x^2}{4(x-1)^4} \ln x+
\frac{-x^3+5 x^2+2 x}{8 (x-1)^3}\nonumber \, \, , \\ 
C_9^{SM}(m_W)&=&-\frac{1}{sin^2\theta_{W}} B(x) +
\frac{1-4 \sin^2 \theta_W}{\sin^2 \theta_W} C(x)-D(x)+\frac{4}{9}, \nonumber \, \, , \\
C_{10}^{SM}(m_W)&=&\frac{1}{\sin^2\theta_W}
(B(x)-C(x))\nonumber \,\, , \\
\label{CoeffW}
\end{eqnarray}
and the primed ones are 
\begin{eqnarray}
C^{\prime SM}_{1,\dots 12}(m_W)&=&0  .
\label{CoeffW2}
\end{eqnarray}
The functions appear in these coefficients are 
\begin{eqnarray}
B(x)&=&\frac{1}{4}\left[\frac{-x}{x-1}+\frac{x}{(x-1)^2} \ln
x\right] \nonumber \,\, , \\
C(x)&=&\frac{x}{4}\left[\frac{x/2-3}{x-1}+\frac{3x/2+1}{(x-1)^2}
       \ln x \right] \nonumber \,\, , \\
D(x)&=&\frac{-19x^3/36+25x^2/36}{(x-1)^3}+
       \frac{-x^4/6+5x^3/3-3x^2+16x/9-4/9}{(x-1)^4}\ln x \,\, ,
\label{BCD}
\end{eqnarray}
and in the coefficients $C_{i}^{(\prime) H}$ (eqs. (\ref{CoeffH}) and 
(\ref{CoeffH2})) are
\begin{eqnarray}
F_{1}(y)&=& \frac{y(7-5y-8y^2)}{72 (y-1)^3}+\frac{y^2 (3y-2)}{12(y-1)^4}
\,\ln y \nonumber  \,\, , \\ 
F_{2}(y)&=& \frac{y(5y-3)}{12 (y-1)^2}+\frac{y(-3y+2)}{6(y-1)^3}\, \ln y 
\nonumber  \,\, ,\\ 
G_{1}(y)&=& \frac{y(-y^2+5y+2)}{24 (y-1)^3}+\frac{-y^2} {4(y-1)^4} \, \ln y
\nonumber  \,\, ,\\ 
G_{2}(y)&=& \frac{y(y-3)}{4 (y-1)^2}+\frac{y} {2(y-1)^3} \, \ln y 
\nonumber\,\, ,\\
H_{1}(y)&=& \frac{1-4 sin^2\theta_W}{sin^2\theta_W}\,\, \frac{x
y}{8}\,\left[ 
\frac{1}{y-1}-\frac{1}{(y-1)^2} \ln y \right]-y \left[\frac{47 y^2-79 y+38}{108
(y-1)^3}-\frac{3 y^3-6 y+4}{18(y-1)^4} \ln y \right] 
\nonumber  \,\, , \\ 
L_{1}(y)&=& \frac{1}{sin^2\theta_W} \,\,\frac{x y}{8}\, \left[-\frac{1}{y-1}+
\frac{1}{(y-1)^2} \ln y \right]
\nonumber  \,\, .\\ 
\label{F1G1}
\end{eqnarray}

\section{The functions which appear in the Wilson coefficients 
$C_9^{eff}$ and $C_9^{\prime eff}$} 
The function which represents the one gluon correction to the matrix 
element $O_9$ is \cite{misiak}
\begin{eqnarray}
\tilde\eta(\hat s) = 1 + \frac{\alpha_{s}(\mu)}{\pi}\, \omega(\hat s)\,\, ,
\label{eta}
\end{eqnarray}
and
\begin{eqnarray} 
\omega(\hat s) &=& - \frac{2}{9} \pi^2 - \frac{4}{3}\mbox{Li}_2(\hat s) 
-\frac{2}{3} \ln {\hat s} \ln(1-{\hat s})-\frac{5+4{\hat s}}{3(1+2{\hat s})}
\ln(1-{\hat s}) - \nonumber \\
& &  \frac{2 {\hat s} (1+{\hat s}) (1-2{\hat s})}
{3(1-{\hat s})^2 (1+2{\hat s})} \ln {\hat s} + \frac{5+9{\hat s}-6{\hat s}^2}{6
(1-{\hat s}) (1+2{\hat s})} \,\, , 
\label{omega}
\end{eqnarray}
$h(z,\hat s)$ arises from the one loop contributions of the
four quark operators $O_1, ... ,O_6$ ($O'_1, ... ,O'_6$)
\begin{eqnarray}
h(z, \hat s) &=& -\frac{8}{9}\ln\frac{m_b}{\mu} - \frac{8}{9}\ln z +
\frac{8}{27} + \frac{4}{9} x \\
& & - \frac{2}{9} (2+x) |1-x|^{1/2} \left\{
\begin{array}{ll}
\left( \ln\left| \frac{\sqrt{1-x} + 1}{\sqrt{1-x} - 1}\right| - 
i\pi \right), &\mbox{for } x \equiv \frac{4z^2}{\hat s} < 1 \nonumber \\
2 \arctan \frac{1}{\sqrt{x-1}}, & \mbox{for } x \equiv \frac
{4z^2}{\hat s} > 1,
\end{array}
\right. \\
h(0, \hat s) &=& \frac{8}{27} -\frac{8}{9} \ln\frac{m_b}{\mu} - 
\frac{4}{9} \ln\hat s + \frac{4}{9} i\pi \,\, , 
\label{hfunc}
\end{eqnarray}
where $z=\frac{m_c}{m_b}$ and $\hat{s}=\frac{q^2}{m_b^2}$.
\section{The form factors for the decay $B\rightarrow K^* l^+ l^-$ }
The structure functions appear in eq. (\ref{matr2}) are 
\begin{eqnarray}
A_{tot}&=& A+A' \nonumber \,\, , \\
B_{1\, tot}&=& B_1+B'_1 \nonumber \,\, , \\
B_{2\, tot}&=& B_2+B'_2 \nonumber \,\, , \\
B_{3\, tot}&=& B_3+B'_3 \nonumber \,\, , \\
C_{tot}&=& C+C' \nonumber \,\, , \\
D_{1\, tot}&=& D_1+D'_1 \nonumber \,\, , \\
D_{2\, tot}&=& D_2+D'_2 \nonumber \,\, , \\
D_{3\, tot}&=& D_3+D'_3  \,\, .
\label{hadpar0}
\end{eqnarray}
Here
\begin{eqnarray}
A &=& -C_9^{eff} \frac{V}{m_B + m_{K^*}} - 4 C_7^{eff} \frac{m_b}{q^2} T_1~,
\nonumber\\ 
B_1 &=& -C_9^{eff} (m_B + m_{K^*}) A_1 - 4 C_7^{eff} \frac{m_b}{q^2} (m_B^2 -
m_{K^*}^2) T_2~,  \nonumber \\ 
B_2 &=& -C_9^{eff} \frac{A_2}{m_B + m_{K^*}} - 4 C_7^{eff} \frac{m_b}{q^2} 
\ga T_2 + \frac{q^2}{m_B^2 - m_{K^*}^2} T_3 \dr~,  \nonumber \\
B_3 &=& -C_9^{eff}\frac{ 2 m_{K^*}}{  q^2}(A_3 - A_0) + 4 C_7 
\frac{m_b}{q^2}T_3~,  \nonumber \\ 
C &=& -C_{10} \frac{V}{m_B + m_{K^*}}~,  \nonumber \\  
D_1 &=& -C_{10} (m_B + m_{K^*}) A_1~,  \nonumber \\     
D_2 &=& -C_{10} \frac{A_2}{m_B + m_{K^*}}~,  \nonumber \\ 
D_3 &=& -C_{10} \frac{2 m_{K^*}}{q^2} (A_3 - A_0)~, \nonumber  \\
\label{hadpar1}
\end{eqnarray}
and 
\begin{eqnarray}
A' &=& -C_9^{\prime eff} \frac{V}{m_B + m_{K^*}} - 4 C_7^{\prime eff} \frac{m_b}{q^2} T_1~,
\nonumber\\ 
B'_1 &=& C_9^{\prime eff} (m_B + m_{K^*}) A_1 + 4 C_7^{\prime eff} \frac{m_b}{q^2} (m_B^2 -
m_{K^*}^2) T_2~,  \nonumber \\ 
B'_2 &=& C_9^{\prime eff} \frac{A_2}{m_B + m_{K^*}} + 4 C_7^{\prime eff} \frac{m_b}{q^2} 
\ga T_2 + \frac{q^2}{m_B^2 - m_{K^*}^2} T_3 \dr~,  \nonumber \\
B'_3 &=& C_9^{\prime eff}\frac{ 2 m_{K^*}}{  q^2}(A_3 - A_0) - 4 C_7^{\prime eff} 
\frac{m_b}{q^2}T_3~,  \nonumber \\ 
C' &=& -C'_{10} \frac{V}{m_B + m_{K^*}}~,  \nonumber \\  
D'_1 &=& C'_{10} (m_B + m_{K^*}) A_1~,  \nonumber \\     
D'_2 &=& C'_{10} \frac{A_2}{m_B + m_{K^*}}~,  \nonumber \\ 
D'_3 &=& C'_{10} \frac{2 m_{K^*}}{q^2} (A_3 - A_0)~, \nonumber  \\
\label{hadpar2}
\end{eqnarray}

We use the $q^2$ dependent expression which is calculated in the framework 
of light-cone QCD sum rules in \cite{braun} to calculate the hadronic 
formfactors $V,~A_1,~A_2,~A_0,~T_1,~T_2$ and $T_3$: 
\begin{eqnarray}
F(q^2)=\frac{F(0)}{1-a_F \frac{q^2}{m_B^2}+b_F (\frac{q^2}{m_B^2})^2}\, ,
\label{formfac}
\end{eqnarray}
where the values of parameters $F(0)$, $a_F$ and $b_F$ are listed in Table 3.

\begin{table}[h]
    \begin{center}
    \begin{tabular}{|c|c|c|c|}
    \hline
    \hline \hline
                &$F(0)$              &       $a_F$ &  $b_F$\\
    \hline \hline    
    $A_1$       &$0.34\pm 0.05$      &       $0.60$&  $-0.023$ \\
    $A_2$       &$0.28\pm 0.04$      &       $1.18$&  $ 0.281$ \\
    $V  $       &$0.46\pm 0.07$      &       $1.55$&  $ 0.575$ \\
    $T_1$       &$0.19\pm 0.03$      &       $1.59$&  $ 0.615$ \\
    $T_2$       &$0.19\pm 0.03$      &       $0.49$&  $-0.241$ \\
    $T_3$       &$0.13\pm 0.02$      &       $1.20$&  $ 0.098$ \\
    \hline
        \end{tabular}
        \end{center}
\caption{The values of parameters existing in eq.(\ref{formfac}) for 
the various form factors of the transition $B\rightarrow K^*$.} 
\label{Table2}
\end{table}
\end{appendix}

\newpage

\end{document}